\documentclass{moriond}

\usepackage{amsmath}

\def\stw{\sin^2\theta_W}
\def\stl{\sin^2\theta^\ell_{\rm eff}}

\def\be{\begin{equation}}
\def\ee{\end{equation}}
\def\bea{\begin{eqnarray}}
\def\eea{\end{eqnarray}}

\newcommand{\Photo}{}

\begin{document}
\vspace*{4cm}
\title{Inclusive and differential W and Z at CMS and ATLAS}

\author{ E. Di Marco }

\address{Istituto Nazionale di Fisica Nucleare, p.le Aldo Moro 2,\\
  Roma 00185, Italy}

\maketitle\abstracts{ Several electroweak precision measurements are
  performed by the ATLAS and CMS collaborations at the LHC. The main
  ones are carried out using Drell-Yan production of single W and Z
  boson.  They regard the measurement of the production cross sections
  of W and Z bosons, the mass of the W boson, and $\sin^2\theta_W$.
  The results of the $\sin^2\theta_W$ measurements have an accuracy of
  approximately twice that reached at LEP and SLD.  Other measurements
  reported are about the Drell-Yan differential production cross
  section.}

\section{Introduction}

The LHC and other hadron colliders measure many parameters of the
Standard Model (SM) and they give very important inputs to the global
SM fits~\cite{Baak:2014ora}.  One relevant SM parameter is the W boson
mass $m_W$, measured at the Tevatron and at the LHC. This can be
compared to the SM prediction obtained using the measurement of the
top quark mass $m_t$, which is precisely measured at the LHC and
Tevatron, and the Higgs boson mass $m_H$, which is precisely measured
at the LHC.  In this respect, a global electroweak fit to SM
parameters yields an indirect estimate of $m_W = 80358 \pm 8$
MeV~\cite{Baak:2014ora}, thus with a precision of $10^{-4}$.

Other SM parameters which are discussed in this presentation are the
measurements of the weak mixing angle ($\stw$) and the measurement of
differential cross sections of the Z boson.  These studies are based
on large samples of $\gamma^*/$Z decays to electrons and muons,
collected during the Run 1 or Run 2 of the LHC.  The measurement of
$\stw$ provides an indirect measurement of $m_W$, through the
relation:
\begin{equation}
\sin^2\theta_W = 1 - \frac{m_W^2}{m_Z^2}
\label{eqn:stw}
\end{equation}
Different experiments use Z boson decays into muon pairs and, whenever
possible, also into electron pairs. The standard selection of Z decays
requires both leptons to have a pseudo-rapidity $\vert\eta\vert<2.4$
(or 2.5), but ATLAS also uses electron pairs with a forward electron
which does not traverse the tracker and is identified and measured in
the calorimeters.

\section{Measurement of $\stw$}

The most precise measurement of $\stw$ at hadron colliders is based on
the forward-backward asymmetry $A_{FB}$ in Drell–Yan~$qq \to
\ell^+\ell^-$ events, where $\ell$ stands for muon ($\mu$) or electron
(e). The forward-backward asymmetry is defined using the angle
$\theta^*$ between the outcoming lepton and the incoming quark in the
Collins–Soper reference frame. A needed ingredient to define
$\theta^*$ is the direction of the incoming quark. At the LHC, which
is a pp collider, the direction of the quark is not know
event-by-event, but can be approximated to be the same as the
longitudinal component of the Z boson momentum because antiquarks
originate from the sea and tend to have lower momentum than quarks
which are predominantly valence quarks. Given the uncertainties in the
quark assignment at the LHC, which are largest for central events, the
observed asymmetry is reduced, compared to the true one and this
effect is referred as dilution.
%Figure~\ref{fig:trueasy} shows the true asymmetry and the diluted asymmetry
%at the LHC in bins of Z boson rapidity.
When the Z boson has no longitudinal momentum, the observed asymmetry
becomes 0 as it becomes impossible to identify the quark
direction. The differential cross section at leading order is:
\begin{equation}
  \frac{d\sigma}{d(\cos\theta^*)}\propto 1 + \cos^2\theta^* + A_4\cos\theta^*
  \label{eqn:diffxsec}
\end{equation}
where $\theta^*$ is the polar angle of the negative lepton in the Collins–Soper frame of the dilepton system
and $A_{FB}$ is defined as:
\begin{equation}
  A_{FB} = \frac{3}{8}A_4 = \frac{\sigma_F - \sigma_B}{\sigma_F + \sigma_B}
  \label{eqn:afb}
\end{equation}
where $\sigma_F$ and $\sigma_B$ are the cross sections for leptons in
the forward and backward hemispheres in terms of $\theta^*$.  The weak
mixing angle, $\stw$, is related to the masses of the W and Z bosons
through the relation in Eq.~\ref{eqn:stw}.
%% %
%% \begin{figure}[!htbp]
%%   \centerline{
%%     \includegraphics[width=0.35\linewidth,draft=false]{Fig1a.pdf}
%%     \includegraphics[width=0.35\linewidth,draft=false]{Fig1b.pdf}
%%   }
%% \caption[]{Left: true AFB for the different flavours of interacting
%%   quarks. Right: AFB after dilution in different ranges of y
%% \label{fig:trueasy}}
%% \end{figure}
%% %
In the improved Born approximation the effective mixing angle, $\stl$ , is
defined, which absorbs some of the higher-order corrections.

\subsection{Measurement of $\stl$ at CMS}
\label{sec:stwcms}
CMS measured $\stl$ , using Z decays into electrons and muons
collected at 8 TeV during Run 1. The CMS measurement is based on a
template fit of the $\cos\theta^*$ distributions.  An event
weighting~\cite{Bodek:2010qg} is used to build $\cos\theta^*$
templates, and weights are used for parton distribution function (PDF)
replicas to also constrain PDFs with data.

\begin{figure}[!htbp]
  \centerline{
    \includegraphics[width=0.45\linewidth,draft=false]{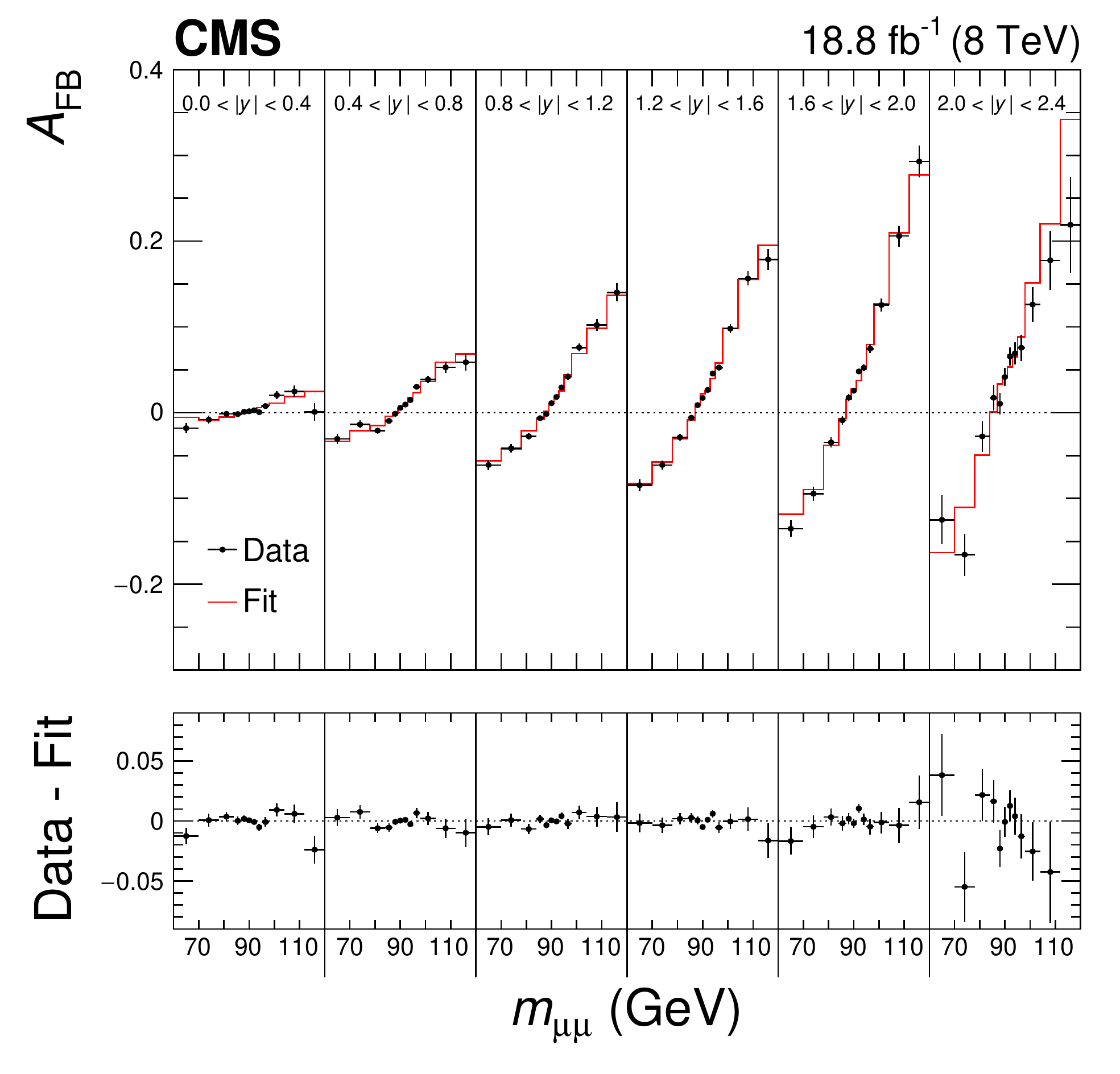}
    \includegraphics[width=0.45\linewidth,draft=false]{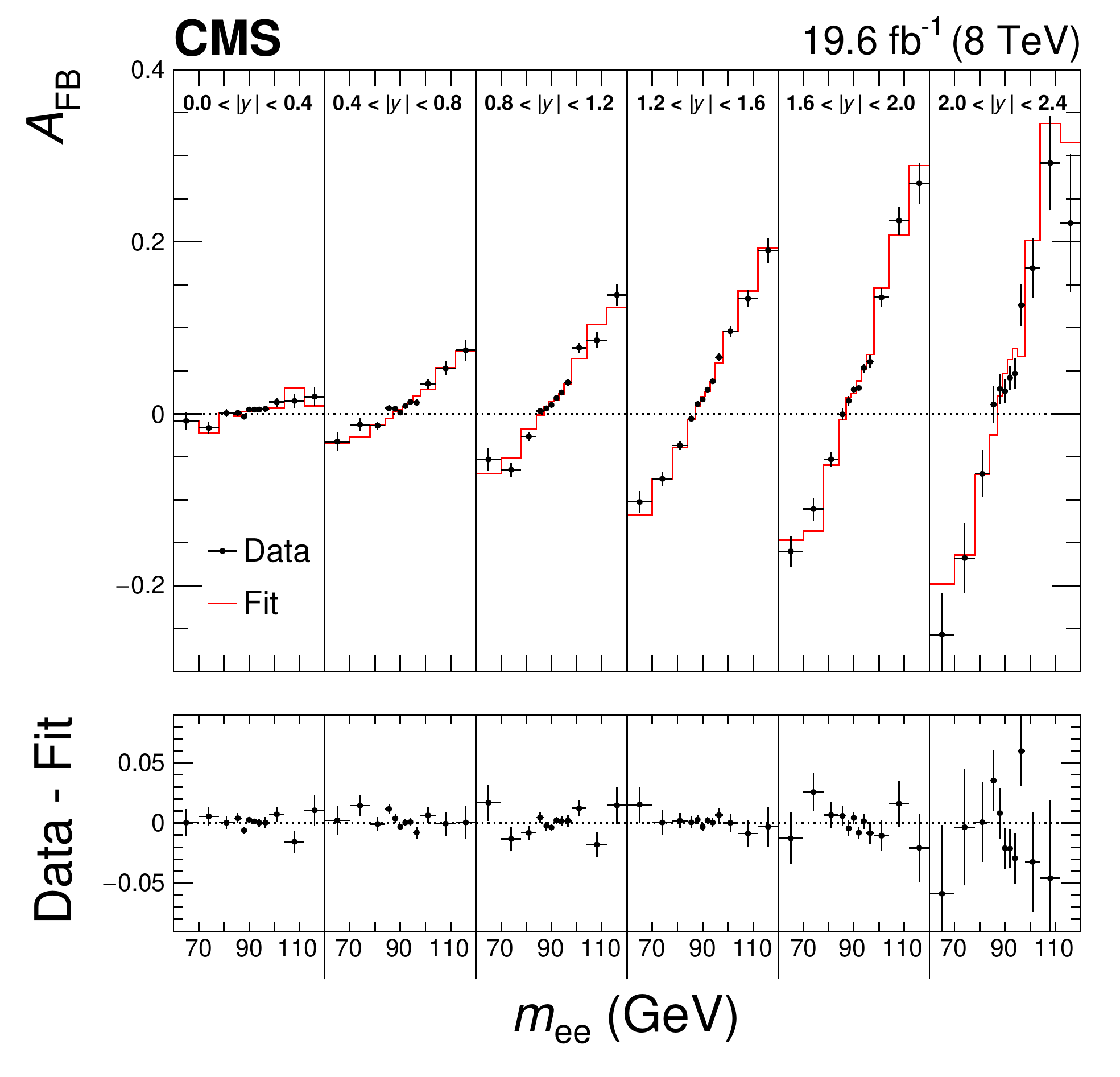}
  }
  \caption[]{
    CMS measurement of $A_{FB}$. Left for Z$\to\mu\mu$. Righ for Z$\to$ee.
    \label{fig:afbcms}}
\end{figure}

The asymmetry measurement is carried out in 6 bins for the dilepton
rapidity $y_{\ell\ell}$ and in 12 bins for the dilepton invariant mass
$m_{\ell\ell}$. The measured data and the results of the fit are shown
in Fig.~\ref{fig:afbcms}. The PDF uncertainties mainly affect AFB far
from the Z boson peak while $\stl$ has the largest effect on the
peak. By eff using weights proportional to $\exp(-\chi^2/2)$, which is
equivalent to profiling over the PDF replicas, the uncertainty due to
PDF decreases from 0.00057 to 0.00030.
The final CMS results~\cite{Sirunyan:2018swq} is:
\begin{equation*}
\stl = 0.23101 \pm 0.00053 = 0.23101 \pm 0.00036(stat) \pm 0.00018(syst) \pm 0.00016(theory) \pm 0.00031(PDF).
\end{equation*}

\subsection{Measurement of $\stl$ and triple differential distributions at ATLAS}
\begin{figure}[!htbp]
  \centerline{
    \includegraphics[width=0.30\linewidth,draft=false]{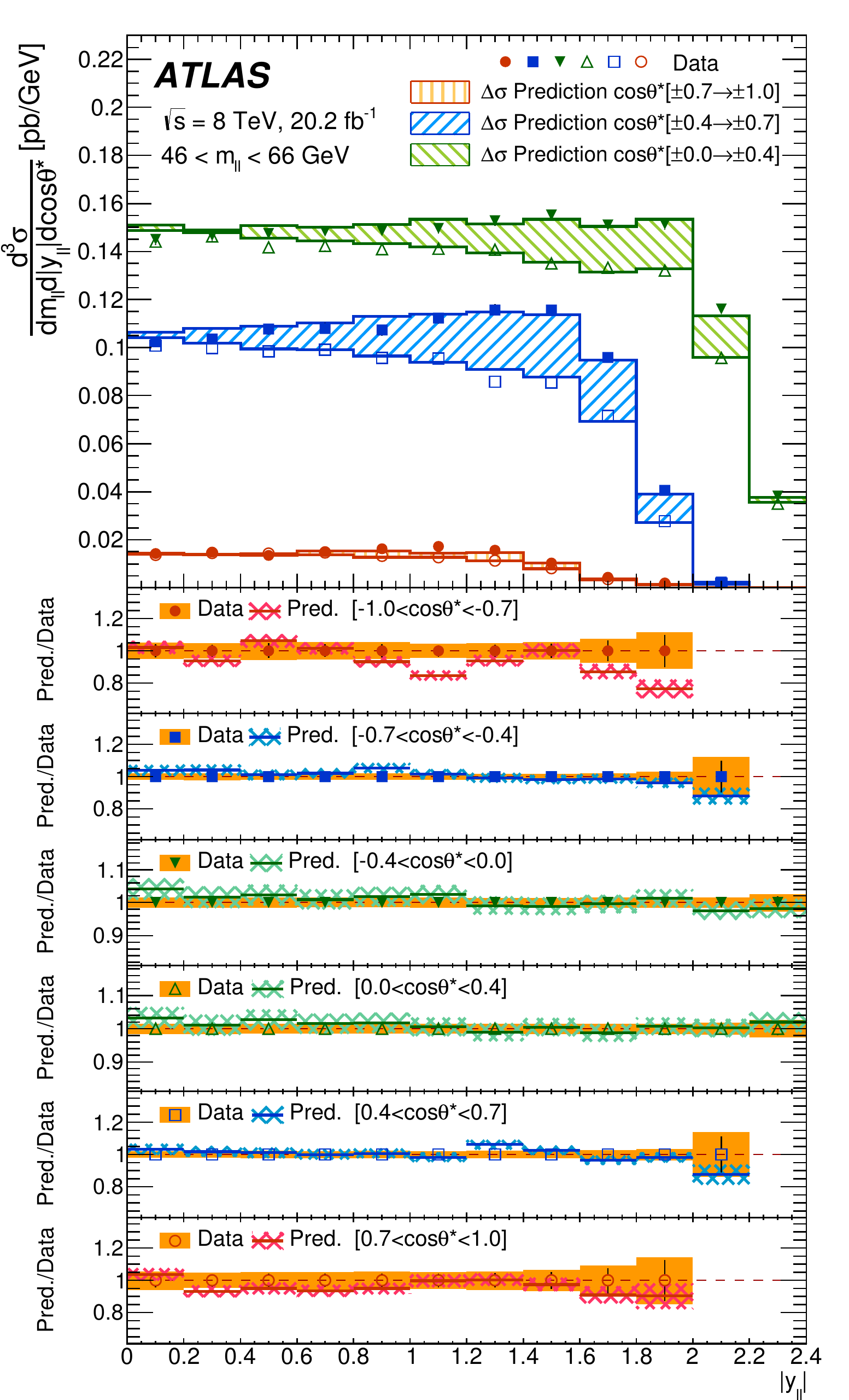}
    \includegraphics[width=0.30\linewidth,draft=false]{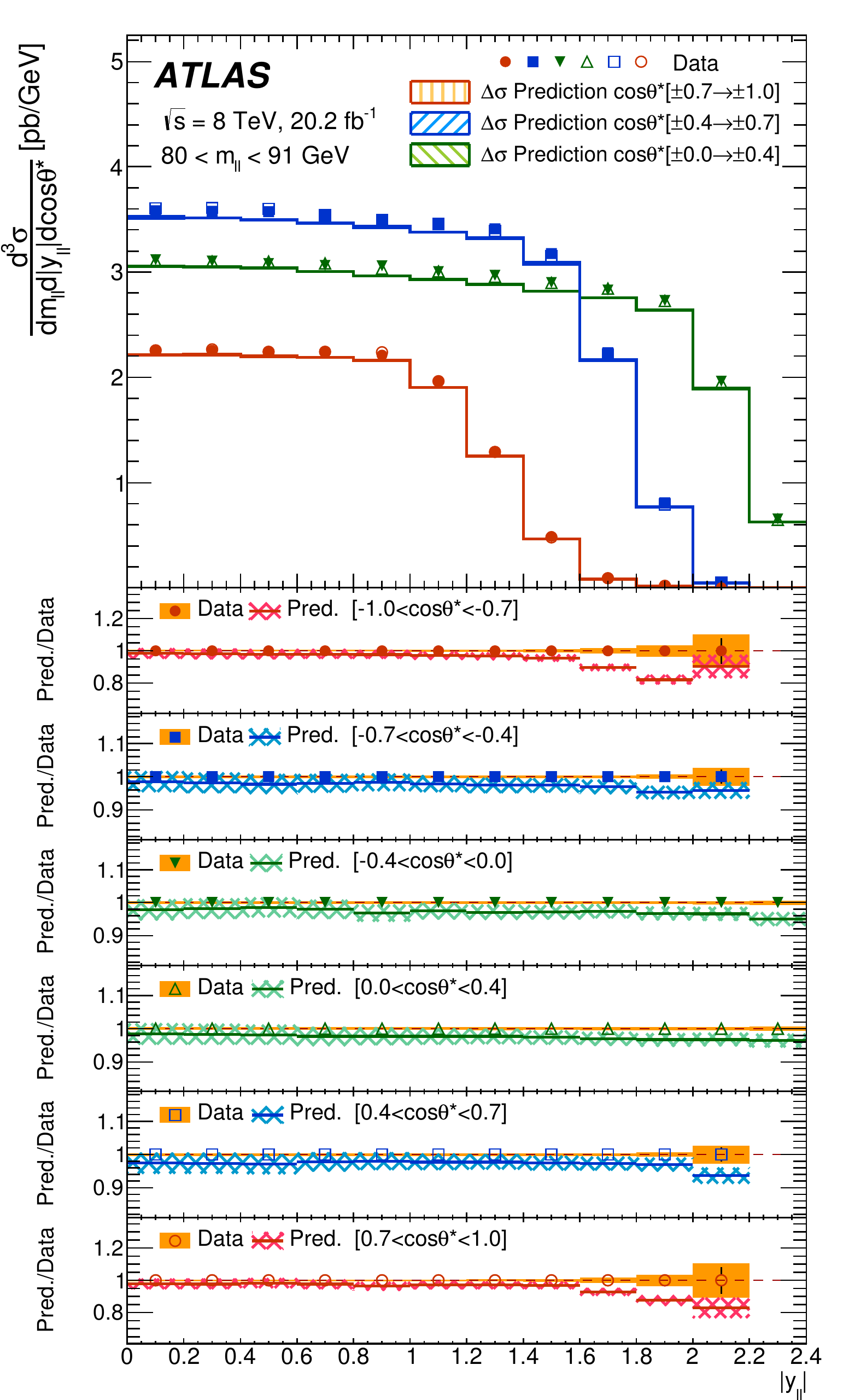}
    \includegraphics[width=0.30\linewidth,draft=false]{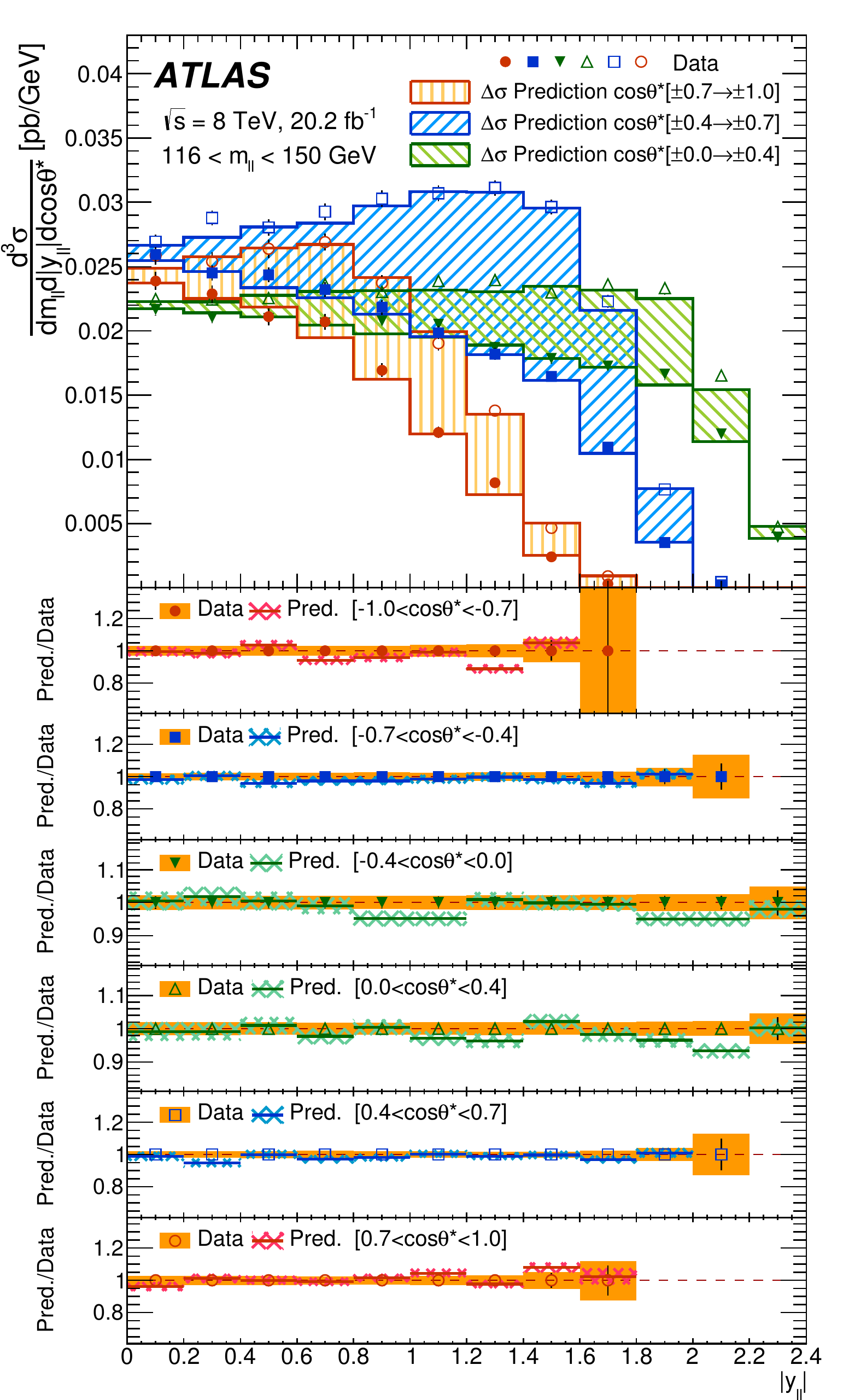}
  }
  \caption[]{
    CMS measurement of AFB. Left for Z$\to\mu\mu$. Righ for Z$\to$ee.
    \label{fig:triplexsec}}
\end{figure}

The ATLAS collaboration previously carried out a measurement of $\stl$
using data collected at 7 TeV. The result is:
\begin{equation*}
\stl = 0.23140 \pm 0.00036 = 0.23140 \pm 0.00021(stat) \pm 0.00016(syst) \pm 0.00024(PDF).
\end{equation*}
The most recent analysis consists in a triple differential cross
section measurement, differential in $m_{\ell\ell}$, $\cos\theta^*$
and $y_{\ell\ell}$~\cite{Aaboud:2017ffb}. The analysis is carried our
using 8 TeV data in the Z$/\gamma\to\mu\mu$ and Z$/\gamma \to$ee
channels, where both leptons are central, and also in Z$/\gamma \to$ee
channel with one central and one forward electron. The analysis is
performed in a total of 504+504+104 bins. Figure~\ref{fig:triplexsec}
shows the measurement as function of $y_{\ell\ell}$ in various
$\cos\theta^*$ bins, positive and negative and for three different
$m_{\ell\ell}$ ranges, below, on and above the Z peak. The data show
good agreement with the predictions of POWHEG. These measurements can
be used to extract the measurement of $\stl$.

\section{Measurement of W boson mass at ATLAS}

The mass of the W boson is measured by ATLAS collaboration using
leptonic decays in 4.6 fb$^{-1}$ of data recorded in 2011 at
$\sqrt{s}=7$ TeV~\cite{Aaboud:2017svj}. Since it is not possible to
fully reconstruct the W boson mass, the measurement relies on
mass-sensitive final state variables, the transverse momentum of the
charged lepton $p_T^\ell$ and the W boson transverse mass $m_T$,
defined as $m_T = \sqrt{2p_T^\ell p_T^{miss}(1-\cos\phi)}$ where
$p_T^{miss}$ is the neutrino missing transverse momentum, estimated
with the vector sum of the transverse energy of all clusters
reconstructed in the calorimeters excluding the lepton
deposits. Templates of these distributions, obtained from simulation
with different values of $m_W$, are compared to the observed
distributions and a $\chi^2$ minimisation is performed to extract the
W boson mass.

Muon momentum scale and resolution corrections
are derived using Z$\to\mu\mu$ (ee) decays.  The combined uncertainty
is dominated by the finite size of the Z boson sample.  The total muon
calibration and efficiency uncertainty in the W boson mass is 9.8 MeV
when estimated using the $p_T^{\ell}$ distribution and 9.7 MeV using
$m_T$.  The hadronic recoil calibration enters the definition of
$m_T$, and it is sensitive to pileup and underlying event
description. Resolution mismodelling is estimated with Z boson events
in data, and transferred to the W boson sample. The uncertainties from
the hadronic recoil calibration affect the W boson mass by 2.6 MeV
when fitting the $p_T^\ell$ distribution and by 13 MeV for $m_T$.

The simulated samples of inclusive vector-boson production are based
on the Powheg MC generator interfaced to Pythia 8, but provide an
imperfect modelling of all the observed distributions. Ancillary
measurements of Drell–Yan processes are used to validate (and tune)
the model and to assess systematic uncertainties. The W and Z boson
samples are reweighted to include the effects of higher-order QCD and
EW corrections, as well as the results of fits to measured
distributions which improve the agreement of the simulated lepton
kinematic distributions with the data. 

The QCD parameters of the parton shower model (PYTHIA 8) were
determined by fits to the transverse momentum distribution of the Z
boson measured at 7 TeV, and it is extrapolated to the W using the
ratio:
\begin{equation}
  R_{W/Z} (p_T) = \left(\frac{1}{\sigma_W}\frac{d\sigma_W(p_T)}{dp_T}\right)
  \left(\frac{1}{\sigma_Z}\frac{d\sigma_Z(p_T)}{dp_T}\right)^{-1}
\label{eqn:wzprop}
\end{equation}
the uncertainties propagated in the ratio are the statistical
precision of the Z data, c-quark and b-quark variations, factorisation
scale variation in the QCD ISR, decorrelating W and Z for the heavy
flavour initiated production while correlating for the light quark
production, leading order PS PDF variations. The uncertainty from the
PDFs on the fixed-order predictions dominate the total uncertainty in
the measurement and amounts to 8.0 MeV in the $p_T^\ell$ fit, and to
8.7 MeV in the $m_T$ fit.

The final combination of all categories gives: $m_W = 80370 \pm 7
(stat.) \pm 11 (exp. syst.) \pm 14 (mod. syst.) MeV$, where the first
uncertainty is statistical, the second corresponds to the experimental
systematic uncertainty, and the third to the physics modelling
systematic uncertainty. The final measurement uncertainty is dominated
by modelling uncertainties, mainly the strong interaction
uncertainties. Lepton calibration uncertainties are the dominant
sources of experimental systematic uncertainty for the extraction of
$m_W$ from the $p_T^\ell$ distribution.
\begin{figure}[!htbp]
  \centerline{
    \includegraphics[width=0.45\linewidth,draft=false]{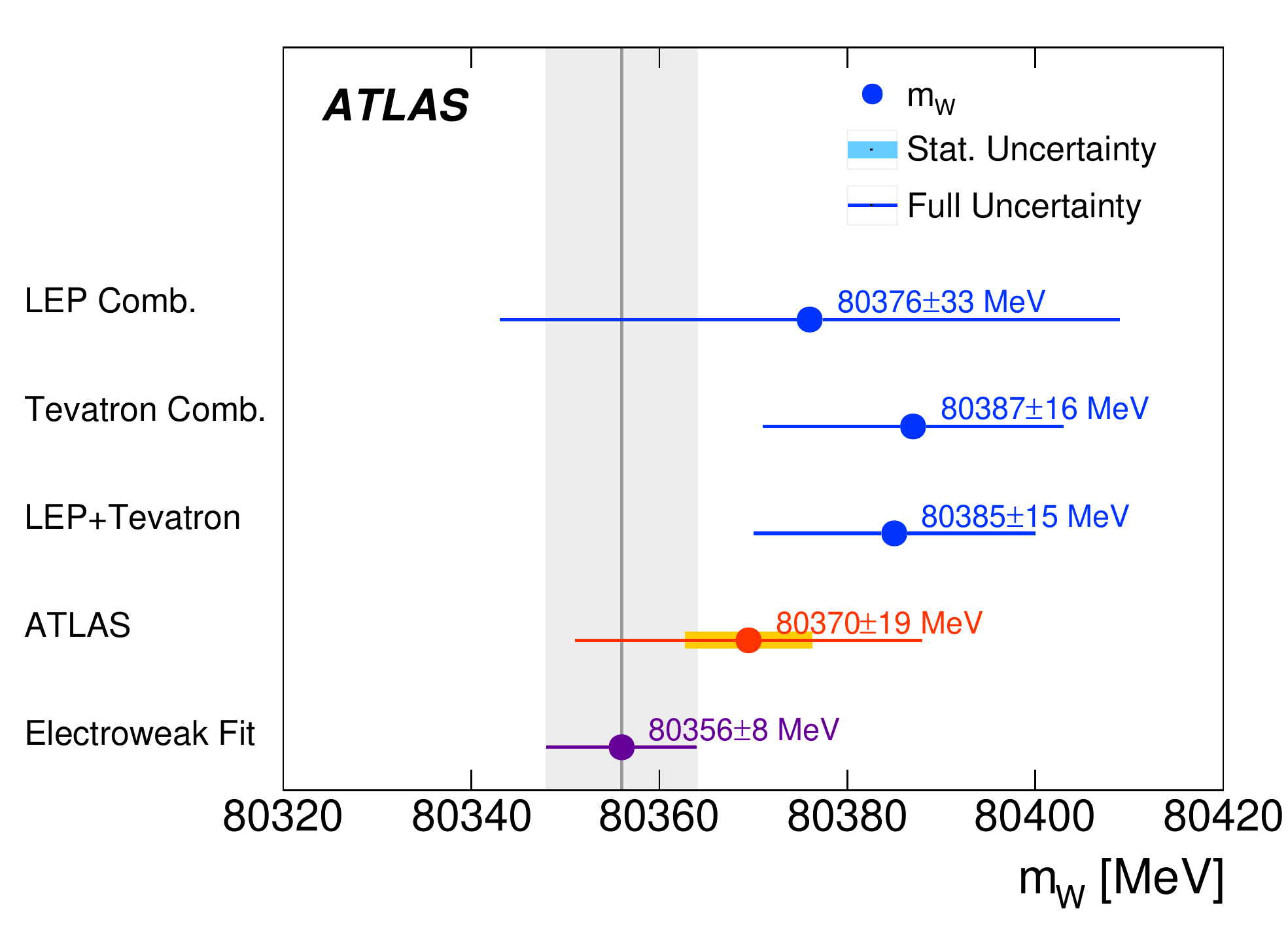}
    \includegraphics[width=0.45\linewidth,draft=false]{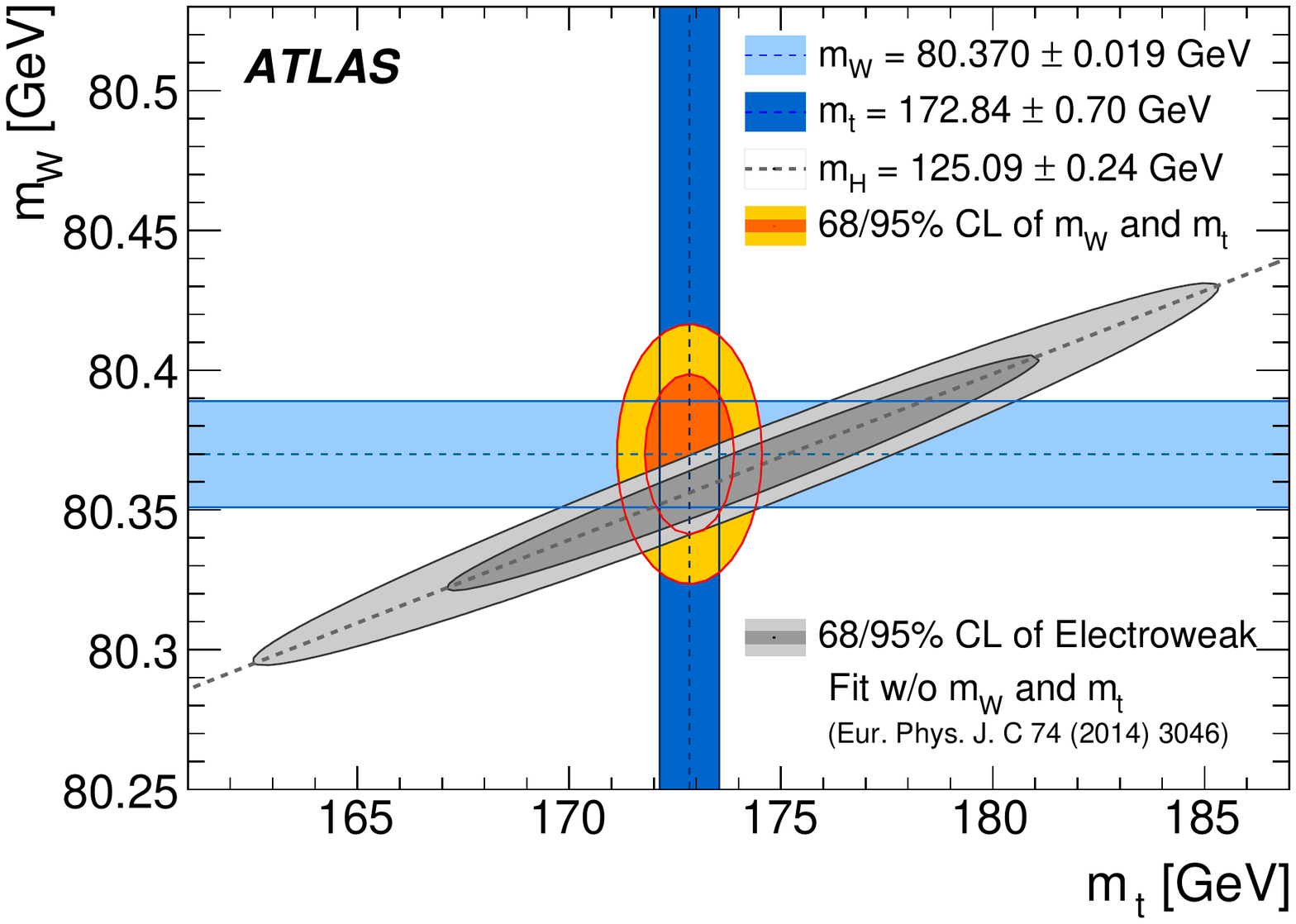}
  }
  \caption[]{ Left: The measured value of mW is compared to other
    published results, including measurements from the LEP
    experiments, and from the Tevatron collider experiments. Right:
    confidence-level contours of the $m_W$ and $m_t$ indirect
    determination from the global electroweak fit are compared to the
    ATLAS measurements of the top-quark and W-boson masses.
    \label{fig:wmass}}
\end{figure}

\end{document}